# Setup for testing LHCb Inner Tracker Modules

P. Vazquez Regueiro [a], D. Esperante Pereira [a], H. Voss [b], L. Nicolas [c]

[a] Universidade de Santiago de Compostela (USC), Spain
[b] Physik-Institut Heidelberg, Germany
[c] Ecole Polytechnique Fédérale Lausanne (EPFL), Switzerland

pablo.vazquez@cern.ch

*Abstract*

The Inner Tracker of the LHCb experiment is a silicon microstrip detector consisting of 336 detector modules with either one or two sensors. The module production is now underway and we present here the setup employed for module testing during the production. The setup is based on the same electronics that will be used in the final experiment. We perform burn-in and ageing tests with the help of a custom made Temperature Cycling Box controlled with LabVIEW under Windows. The DAQ is done in another PC running Linux. Here we integrate the different C/C++ libraries used to communicate to the LHCb Time and Fast Control system, Experiment Control System and Data Acquisition.

## I. THE LHCB EXPERIMENT AND THE INNER TRACKER

The LHCb experiment [1] is one of the four large experiments at the new Large Hadron Collider (LHC) at CERN. It is a single-arm forward spectrometer, with an acceptance from 15 to 300 mrad, dedicated to B-physics. The tracking system is formed by the Vertex detector (VELO), the Trigger Tracker (TT) station and the Tracking stations (T1-T3) behind the magnet which are split into the Inner Tracker (IT) and the Outer Tracker (OT).

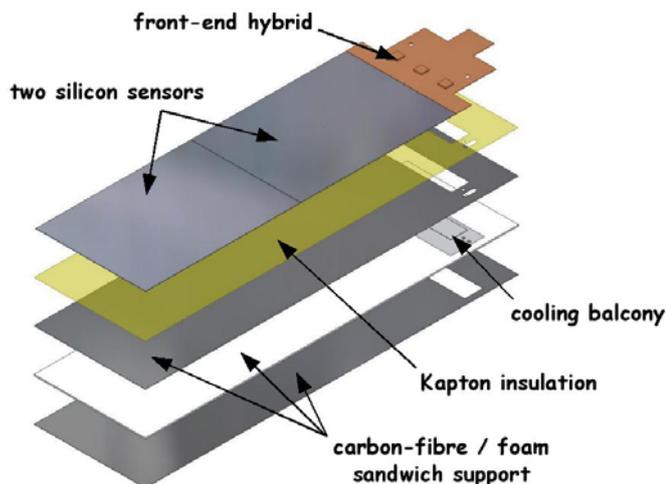

Figure 1: Inner Tracker module composition

The Inner Tracker [2] is a silicon detector covering the high multiplicity region around the beam pipe. With an active area of 4.3 m$^2$ (1.3 % of total acceptance area of LHCb) and 129,024 channels, it covers the 20% of the tracks. Each station consists of four boxes, and 4 layers of seven modules per box. The boxes on top and bottom of the beam pipe have 1-sensor modules of 320µm thickness while the boxes at the right and left have 2-sensor modules (410µm). The silicon sensors are single sided 108 mm long, 384 AC coupled readout strips, a pitch of 198 µm and a width over pitch ratio w/p = 0.25. The three very front-end readout chips, called Beetles [3], are mounted on an electronics hybrid. A high thermal conductive carbon fibre is used as a support of the sensor/s, the ceramic pitch adapter with gold strips, the hybrid and an aluminium piece "balcony" for positioning and cooling, as shown in Figure 1.

## II. TESTING HARDWARE

We use for the module testing a custom made temperature cycling box with temperature and humidity control. It provides a capacity for 6 modules in a volume of 25x30x50 cm$^3$. The box and its control electronics were fully designed and built in Santiago. It is based on water cooled Peltier elements and allows to maintain fixed temperatures from +50 ºC to -15 ºC with an accurancy of 1 ºC.

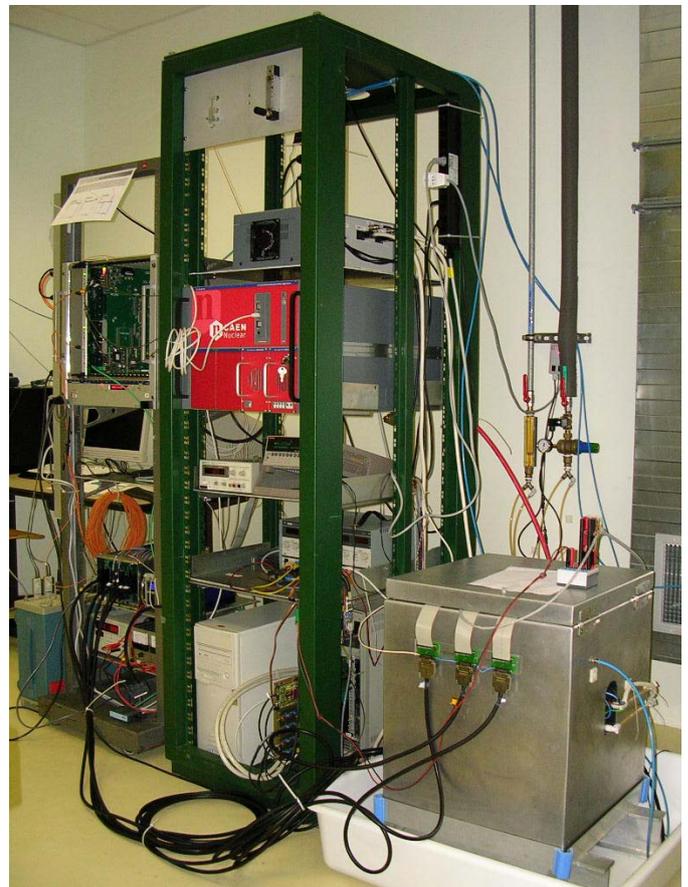

Figure 2: IT module burn-in and temperature cycling setup

The DAQ setup used in the testing uses the same electronics as the final experiment. The frontend readout electronics: Control Boards and Digitizer Boards, the clock and trigger distribution: Readout Supervisor [4], the common LHCb readout board: TELL1 [5], high voltage: CAEN system with module A1511B. The IV-curves are done using a Keithley picoampermeter and the metrology using a XY table with a Leica microscope. Temperatue cycling box and DAQ setup can be seen in Figure 2.

## III. TESTING PROCEDURES

We produce the modules in two sites [6]. In Lausanne the EPFL group glue the sensor, pitch adapter, hybrid and balcony into the support. In the clean room at CERN the USC/EPFL group do the bonding, testing and metrology as well as the final mounting of the modules in the boxes.

Once the modules arrive to CERN we do IV-curves before and after the bonding, then we perform the temperature cycling/burn-in over 44 h, 30 temperature cycles from -5 ºC to 40 ºC. A temperature cycle involves 8 temperature points: 20 ºC, 40 ºC, 20 ºC, 10 ºC, 0 ºC, -5 ºC, 0 ºC, 10 ºC. In order to test modules at different temperatures we take data every 9 temperature points, once per cycle. All hybrids have passed a previous burn-in test in Heidelberg before module assembly. In every data taking we test for open/shorted channels by injecting charge via the Beetle test-pulse in 30K events. After the burn-in we test for pinholes, then we do the metrology of the modules and store them in an inert atmosphere until they are mounted in detector boxes. In Figure 3 the failure ratio for the production chain can be seen.

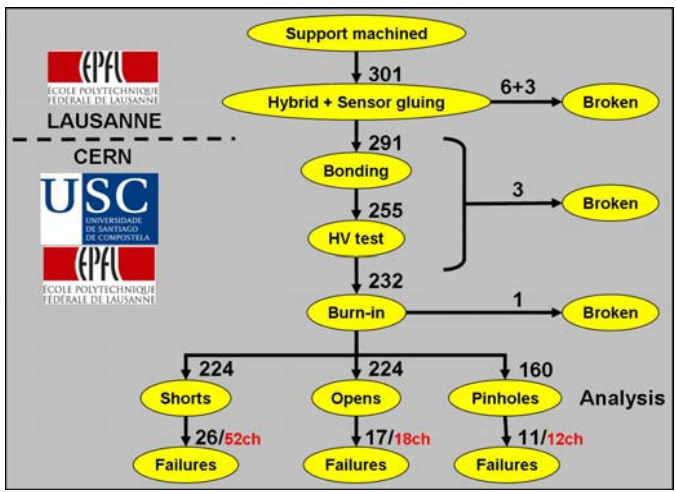

Figure 3: IT module production scheme by August 2006

## IV. TESTING SOFTWARE

The control of the testing setup is based on LabVIEW. A PC running windows controls the Temperature Cycling Box through the parallel port while another PC running Linux acquires data. The high voltage is controlled via an OPC server. C++ based readout software allows to perform different actions as ADC timing scan, pedestal acquisition, test pulse delay scan. We are forced to test either short or long modules in the same burn-in as they have different pulse timing due to the different input capacity. Some software and firmware adaptations were needed in order to run the final LHCb trigger, control and readout electronics.

## V. ANALYSIS

With the data collected during the burn-in we characterize modules and look for bad channels [7]. Raw and common mode suppressed noise are monitored for different temperatures. We also perform a delay scan to determine the pulse-shape of every channel at different temperatures. Shorted channels are identified by smaller signal heights due to larger load capacitance caused by two strips connected to the front-end channel rather than just one. Unbonded or open channels are identified by larger signal heights due to smaller load capacitance. Pinholes are identified in a different test by artificial low noise due to preamplifier saturation under infrared illumination at room temperature. Figure 4 shows a delay scan and shortcut channels for module 121.

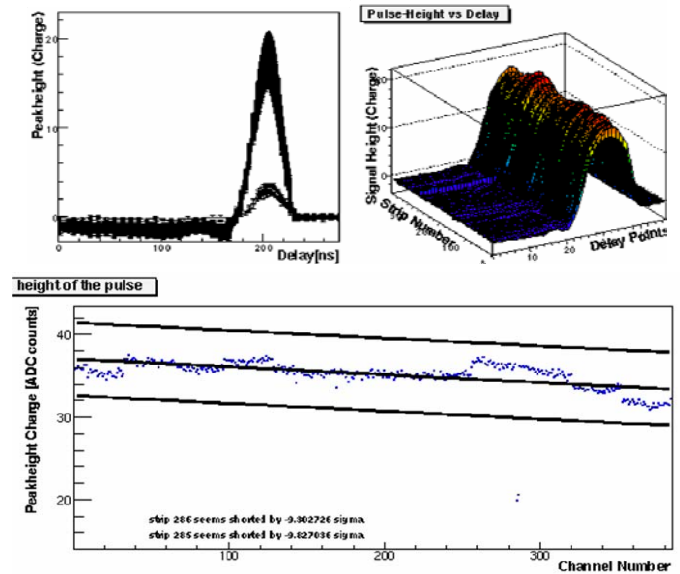

Figure 4: Delay scan and open/short analysis for a IT module

## VI. BONDING

We use two wedge bonding machines to perform the module bonding, a Kulicke and Soffa 8060 and a Delvotec 6319. Both use 25 µm diameter aluminium wire. We have encountered large variations in the bondability of pitch adapters, so we have to check and possibly adapt the bond parameters of each individual module by pulling some test bonds placed on strips of the pitch adapter. For this we use a DAGE 3000 pull tester. Figure 5 shows the detail of the bonding of a Beetle chip.

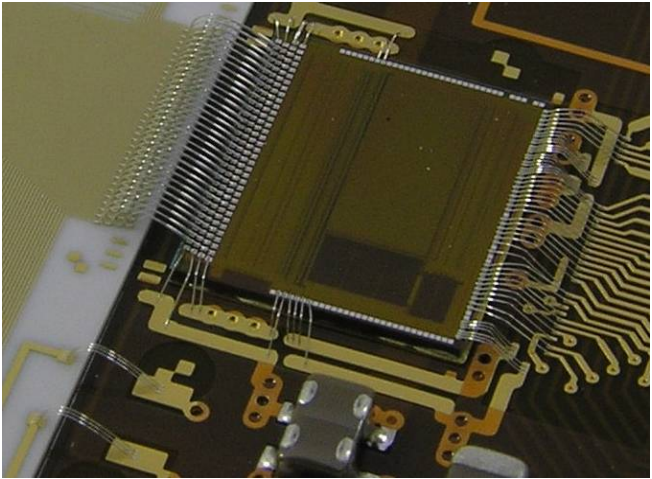

Figure 5: Beetle to pitch adapter bonding in 4 rows of 40 µm pitch

## VII. BOX ASSEMBLY

We have assembled the first IT box and we have tested the interface PCB used to interconnect the modules to the external world and to distribute the high voltage lines. The 6 carbon fibre support frames are complete and ready for cabling and installation in the pit. Half of the detector boxes are already produced in Santiago and delivered to CERN. The production of the other half is well advanced.

## VIII. SUMMARY

We have presented the setup employed in the testing of LHCb Inner Tracker modules and testing procedures. Using this setup we have tested so far about 250 detector modules and encountered one electronic failure of a module during the temperature cycling. The testing is running smoothly and overall production is progressing well